\documentclass[english]{article}
\usepackage[T1]{fontenc}
\usepackage[latin9]{inputenc}
\usepackage{color}
\usepackage{textcomp}
\usepackage{amssymb}
\usepackage{babel}
\begin{document}

\title{\textbf{Quantum transitions of minimum energy for Hawking quanta
in highly excited black holes: problems for loop quantum gravity?}}

\author{\textbf{Christian Corda}}

\maketitle
\begin{center}
Institute for Theoretical Physics and Advanced Mathematics (IFM) Einstein-Galilei,
Via Santa Gonda 14 - 59100 Prato, Italy
\par\end{center}

\begin{center}
Istituto Universitario di Ricerca \textquotedbl{}Santa Rita\textquotedbl{},
Villa il Ventaglio (G.C.) via delle Forbici 24/26 - 50133 Firenze,
Italy
\par\end{center}

\begin{center}
International Institute for Applicable Mathematics and Information
Sciences (IIAMIS), Hyderabad, India and Udine, Italy
\par\end{center}

\begin{center}
\textit{E-mail address:} \textcolor{blue}{cordac.galilei@gmail.com} 
\par\end{center}
\begin{abstract}
By analysing some recent results by Yoon, which arise from loop quantum
gravity and from the assumption of the locality of photon emission
in a black hole, we argue that they are not consistent with our recent
semi-classical results for highly excited black holes. Maybe that
the results by Yoon can be correct for non-highly excited black holes,
but, in any case, our analysis renders further problematical the match
between loop quantum gravity and semi-classical theory.
\end{abstract}
It has been suggested \cite{key-1} that the Hawking radiation spectrum
\cite{key-2} could be discrete if one quantized the area spectrum
in a way that the allowed area is the integer multiples of a single
unit area. Indeed, the Hawking radiation spectrum looks to be continuous
in loop quantum gravity if the area spectrum is quantized in such
a way that there are not only a single unit area \cite{key-3,key-4}. 

Recently, by assuming the locality of photon emission in a black hole,
Yoon argued that the Hawking radiation spectrum is discrete in the
framework of loop quantum gravity even in the case that the allowed
area is not simply the integer multiples of a single unit area \cite{key-5}.
Yoon's result arises from the selection rule for quantum black holes
\cite{key-5}.

On the other hand, by analysing Hawking radiation as tunnelling, Parikh
and Wilczek showed that the radiation spectrum cannot be strictly
thermal \cite{key-6,key-7}. In fact, the energy conservation implies
that the black hole contracts during the process of radiation\cite{key-6,key-7}.
Thus, the horizon recedes from its original radius to a new, smaller
radius \cite{key-6,key-7}. The consequence is that black holes cannot
strictly emit thermally \cite{key-6,key-7}. This is consistent with
unitarity \cite{key-6} and has profound implications for the black
hole information puzzle because arguments that information is lost
during black hole's evaporation rely in part on the assumption of
strict thermal behavior of the spectrum \cite{key-6,key-7,key-8}.

\noindent Working with $G=c=k_{B}=\hbar=\frac{1}{4\pi\epsilon_{0}}=1$
(Planck units), the probability of emission is \cite{key-2,key-6,key-7}
\begin{equation}
\Gamma\sim\exp(-\frac{\omega}{T_{H}}),\label{eq: hawking probability}
\end{equation}

\noindent where $T_{H}\equiv\frac{1}{8\pi M}$ is the Hawking temperature
and $\omega$ the energy-frequency of the emitted radiation.

\noindent Parikh and Wilczek released a remarkable correction, due
to an exact calculation of the action for a tunnelling spherically
symmetric particle, which yields \cite{key-6,key-7}

\noindent 
\begin{equation}
\Gamma\sim\exp[-\frac{\omega}{T_{H}}(1-\frac{\omega}{2M})].\label{eq: Parikh Correction}
\end{equation}

\noindent This important result, which takes into account the conservation
of energy, enables a correction, the additional term $\frac{\omega}{2M}$
\cite{key-6,key-7}. We recently finalized the tunneling framework
by Parikh and Wilczek in \cite{key-20}, by showing that the probability
of emission (\ref{eq: Parikh Correction}) is correctly associated
to the two distributions \cite{key-20} 
\begin{equation}
<n>_{boson}=\frac{1}{\exp\left[-4\pi n\left(M-\omega\right)\omega\right]-1},\;\;<n>_{fermion}=\frac{1}{\exp\left[-4\pi n\left(M-\omega\right)\omega\right]+1},\label{eq: final distributions}
\end{equation}
for bosons and fermions respectively, which are \emph{non} strictly
thermal. 

\noindent In various frameworks of physics and astrophysics the deviation
from the thermal spectrum of an emitting body is taken into account
by introducing an \emph{effective temperature }which represents the
temperature of a black body that would emit the same total amount
of radiation \cite{key-9}. The effective temperature can be introduced
for black holes too \cite{key-9}. It depends on the energy-frequency
of the emitted radiation and is defined as \cite{key-9}

\noindent 
\begin{equation}
T_{E}(\omega)\equiv\frac{2M}{2M-\omega}T_{H}=\frac{1}{4\pi(2M-\omega)}.\label{eq: Corda Temperature}
\end{equation}

\noindent Then, eq. (\ref{eq: Parikh Correction}) can be rewritten
in Boltzmann-like form \cite{key-9}

\noindent 
\begin{equation}
\Gamma\sim\exp[-\beta_{E}(\omega)\omega]=\exp(-\frac{\omega}{T_{E}(\omega)}),\label{eq: Corda Probability}
\end{equation}

\noindent where $\beta_{E}(\omega)\equiv\frac{1}{T_{E}(\omega)}$
and $\exp[-\beta_{E}(\omega)\omega]$ is the \emph{effective Boltzmann
factor} appropriate for an object with inverse effective temperature
$T_{E}(\omega)$ \cite{key-9}. The ratio $\frac{T_{E}(\omega)}{T_{H}}=\frac{2M}{2M-\omega}$
represents the deviation of the radiation spectrum of a black hole
from the strictly thermal feature \cite{key-9}. If $M$ is the initial
mass of the black hole \emph{before} the emission, and $M-\omega$
is the final mass of the hole \emph{after} the emission \cite{key-9},
eqs. (\ref{eq: Parikh Correction}) and (\ref{eq: Corda Temperature})
enable the introduction of the \emph{effective mass }and of the \emph{effective
horizon} \cite{key-9} 
\begin{equation}
M_{E}\equiv M-\frac{\omega}{2},\mbox{ }r_{E}\equiv2M_{E}\label{eq: effective quantities}
\end{equation}

\noindent of the black hole \emph{during} the emission of the particle,
i.e. \emph{during} the contraction's phase of the black hole \cite{key-9}.
The \emph{effective quantities $T_{E},$ $M_{E}$ }and\emph{ $r_{E}$
}are average quantities. \emph{$M_{E}$ }is the average of the initial
and final masses, \emph{$r_{E}$ }is the average of the initial and
final horizons and \emph{$T_{E}$ }is the inverse of the average value
of the inverses of the initial and final Hawking temperatures (\emph{before}
the emission $T_{H\mbox{ initial}}=\frac{1}{8\pi M}$, \emph{after}
the emission $T_{H\mbox{ final}}=\frac{1}{8\pi(M-\omega)}$) \cite{key-9}.
Notice that the analysed process is \emph{discrete} instead of \emph{continuous}
\cite{key-9}. In fact, the black hole's state before the emission
of the particle and the black hole's state after the emission of the
particle are different countable black hole's physical states separated
by an \emph{effective state} which is characterized by the effective
quantities \cite{key-9}. Hence, the emission of the particle can
be interpreted like a \emph{quantum} \emph{transition} of frequency
$\omega$ between the two discrete states \cite{key-9}. The tunnelling
visualization is that whenever a tunnelling event works, two separated
classical turning points are joined by a trajectory in imaginary or
complex time \cite{key-6,key-9}. 

In \cite{key-9} we used the concepts of effective quantities and
the discrete character of Hawking radiation to argue a natural correspondence
between Hawking radiation and black hole's quasi-normal modes. A problem
concerning previous attempts to associate quasi-normal modes to Hawking
radiation was that ideas on the continuous character of Hawking radiation
did not agree with attempts to interpret the frequency of the quasi-normal
modes \cite{key-10}. In fact, the discrete character of the energy
spectrum of black hole's quasi-normal modes should be incompatible
with the spectrum of Hawking radiation whose energies are of the same
order but continuous \cite{key-10}. Actually, the issue that Hawking
radiation is not strictly thermal and, as we have shown, it has discrete
instead of continuous character, removes the above difficulty \cite{key-9}.
In other words, the discrete character of Hawking radiation permits
to interpret black hole's quasi-normal frequencies in terms of energies
of physical Hawking quanta too \cite{key-9}. In fact, quasi-normal
modes are damped oscillations representing the reaction of a black
hole to small, discrete perturbations \cite{key-9,key-11,key-12,key-13}.
A discrete perturbation can be the capture of a particle which causes
an increase in the horizon area \cite{key-11,key-12,key-13}. Hence,
if the emission of a particle which causes a decrease in the horizon
area is a discrete rather than continuous process, it is quite natural
to assume that it is also a perturbation which generates a reaction
in terms of countable quasi-normal modes \cite{key-9}. This natural
correspondence between Hawking radiation and black hole's quasi-normal
modes permits to consider quasi-normal modes in terms of quantum levels
not only for absorbed energies like in \cite{key-11,key-12,key-13},
but also for emitted energies like in \cite{key-9}. This issue endorses
the idea that, in an underlying unitary quantum gravity theory, black
holes can be considered highly excited states \cite{key-9,key-11,key-12,key-13}
and looks consistent with Yoon's approach in loop quantum gravity
\cite{key-5}.

\noindent The intriguing idea that black hole's quasi-normal modes
carry important information about black hole's area quantization is
due to the remarkable works by Hod \cite{key-11,key-12}. Hod's original
proposal found various objections over the years \cite{key-9,key-13}
which have been answered in a good way by Maggiore \cite{key-13},
who refined Hod's conjecture. In \cite{key-9} we further improve
the Hod-Maggiore conjecture by taking into account the non-strict
thermal character and, in turn, discrete rather than continuous character
of Hawking radiation spectrum. In particular, we found the solution
for the absolute value of the quasi-normal frequencies in the case
of large $n$ (highly excited black hole) \cite{key-9} 
\begin{equation}
(\omega_{0})_{n}=M-\sqrt{M^{2}-\frac{1}{4\pi}\sqrt{(\ln3)^{2}+4\pi^{2}(n+\frac{1}{2})^{2}}},\label{eq: radice fisica}
\end{equation}
where $n$ is the \emph{quantum ``overtone''number}, see \cite{key-9}
for details. Eq. (\ref{eq: radice fisica}) is based on Eq. (7) in
\cite{key-9}, which is 
\begin{equation}
\begin{array}{c}
\omega_{n}=\ln3\times T_{E}(|\omega_{n}|)+2\pi i(n+\frac{1}{2})\times T_{E}(|\omega_{n}|)+\mathcal{O}(n^{-\frac{1}{2}})=\\
\\
=\frac{\ln3}{4\pi(2M-|\omega_{n}|)}+\frac{2\pi i}{4\pi(2M-|\omega_{n}|)}(n+\frac{1}{2})+\mathcal{O}(n^{-\frac{1}{2}}).
\end{array}\label{eq: quasinormal modes corrected}
\end{equation}
We intuitively derived eq. (\ref{eq: quasinormal modes corrected})
in \cite{key-9}. A rigorous analytical derivation of it can be found
in the Appendix of \cite{key-21}.

We also found that, for large $n,$ an emission involving the levels
$n$ and $n-1$  of a Schwarzschild black hole having an original
mass M gives a variation of energy \cite{key-9}

\noindent 
\begin{equation}
\begin{array}{c}
\Delta E=(\omega_{0})_{n}-(\omega_{0})_{n-1}=f(M,n)\equiv\\
\\
\equiv\sqrt{M^{2}-\frac{1}{4\pi}\sqrt{(\ln3)^{2}+4\pi^{2}(n-\frac{1}{2})^{2}}}-\sqrt{M^{2}-\frac{1}{4\pi}\sqrt{(\ln3)^{2}+4\pi^{2}(n+\frac{1}{2})^{2}}}.
\end{array}\label{eq: f(M,n)}
\end{equation}
The result of eq. (\ref{eq: f(M,n)}) also hold for a Kerr black hole
in the case $M^{2}\gg J,$ where $J$ is the angular momentum of the
black hole \cite{key-22}.The analysis in \cite{key-5} shows that
the Hawking radiation spectrum is truncated below a certain frequency
and hence there is a minimum energy of an emitted particle \cite{key-5}

\begin{equation}
E_{min}\approx\alpha T_{H}=\frac{\alpha}{8\pi M},\label{eq: energia minima}
\end{equation}
where $\alpha=1.49$ in the case of isolated horizon framework \cite{key-14,key-15},
$\alpha=2.46$ in the case of the Tanaka-Tamaki scenario \cite{key-16}
and $\alpha=4.44$ in the case of the Kong-Yoon scenario \cite{key-17,key-18,key-19}.

Hence, by considering an excited black hole, we argue that they should
exist values of $n$ that we label as $n_{*}$ for which 

\begin{equation}
f(M,n_{*})=E_{min}.\label{eq: n minimo}
\end{equation}
In other words, we search the minimum energy of an emitted particle
for a fixed Hawking temperature which corresponds to two neighboring
levels. In fact, if the two levels are not neighboring the emitted
energy will be higher. Thus, $n_{*}$ will be the values of $n$ for
which Hawking quanta having minimum energy can be emitted in emissions
involving two neighboring levels ($n_{*}$ and $n_{*}-1$). 

A black hole excited at a level $n_{*}-1$ has a mass \cite{key-9}

\begin{equation}
M_{n_{*}-1}\equiv M-(\omega_{0})_{n-1}\label{eq: M n-1}
\end{equation}

which, by using eq. (\ref{eq: radice fisica}) becomes

\begin{equation}
M_{n_{*}-1}=\sqrt{M^{2}-\frac{1}{4\pi}\sqrt{(\ln3)^{2}+4\pi^{2}(n_{*}-\frac{1}{2})^{2}}}\label{eq: M n-1 a}
\end{equation}
Considering eqs. (\ref{eq: f(M,n)}), (\ref{eq: energia minima})
and (\ref{eq: n minimo}) one gets

\begin{equation}
\begin{array}{c}
\sqrt{M^{2}-\frac{1}{4\pi}\sqrt{(\ln3)^{2}+4\pi^{2}(n_{*}-\frac{1}{2})^{2}}}\approx\\
\\
\approx\sqrt{M^{2}-\frac{1}{4\pi}\sqrt{(\ln3)^{2}+4\pi^{2}(n_{*}+\frac{1}{2})^{2}}}+\frac{\alpha}{8\pi\sqrt{M^{2}-\frac{1}{4\pi}\sqrt{(\ln3)^{2}+4\pi^{2}(n_{*}-\frac{1}{2})^{2}}}}.
\end{array}\label{eq: bilanciamento}
\end{equation}
For $n\gg1$ eq. (\ref{eq: bilanciamento}) is well approximated by
\begin{equation}
\begin{array}{c}
\sqrt{M^{2}-\frac{1}{2}n_{*}+\frac{1}{4}}\approx\\
\\
\approx\sqrt{M^{2}-\frac{1}{2}n_{*}-\frac{1}{4}}+\frac{\alpha}{8\pi\sqrt{M^{2}-\frac{1}{2}n_{*}+\frac{1}{4}}}.
\end{array}\label{eq: bilanciamento-1}
\end{equation}
Putting 
\begin{equation}
\begin{array}{c}
M^{2}-\frac{1}{2}n_{*}+\frac{1}{4}\equiv x\\
\\
\frac{\alpha}{8\pi}\equiv\beta
\end{array}\label{eq: putting}
\end{equation}
eq. (\ref{eq: bilanciamento-1}) becomes

\begin{equation}
\sqrt{x}\approx\sqrt{x-\frac{1}{2}}+\frac{\beta}{\sqrt{x}}\label{eq: bilanciamo}
\end{equation}
which can be simplified as

\begin{equation}
x-\beta\approx\sqrt{x-\frac{1}{2}}\sqrt{x}.\label{eq: bilanciamo 2}
\end{equation}
By squaring eq. (\ref{eq: bilanciamo 2}) one easily solve for $x$:

\begin{equation}
x\approx\frac{2\beta^{2}}{4\beta-1}.\label{eq: x}
\end{equation}
Hence, by using eq. (\ref{eq: putting}) one gets immediately

\begin{equation}
n_{*}\approx2M^{2}+\frac{1}{2}-\frac{\alpha^{2}}{8\pi\left(\alpha-2\pi\right)}.\label{eq: n star}
\end{equation}
Thus, we find a sole value of $n_{*}$ for which the Hawking radiation
spectrum is truncated below a certain frequency of minimum energy
for a transition between two neighboring levels $n_{*}$ and $n_{*}-1$
of an highly excited black hole. By using eqs. (\ref{eq: n star})
and (\ref{eq: M n-1 a}) the black hole's mass at the level $n_{*}-1$
results

\begin{equation}
\begin{array}{c}
M_{n_{*}-1}=\sqrt{M^{2}-\frac{1}{4\pi}\sqrt{(\ln3)^{2}+4\pi^{2}(n_{*}-\frac{1}{2})^{2}}}\approx\\
\\
\sqrt{M^{2}-\frac{1}{2}n_{*}+\frac{1}{4}}\approx\sqrt{\frac{\alpha^{2}}{16\pi\left(\alpha-2\pi\right)}}
\end{array}\label{eq: M n-1 a-1}
\end{equation}
We note that the mass (\ref{eq: M n-1 a-1}) and its correspondent
Hawking temperature 

\begin{equation}
\left(T_{H}\right)_{n_{*}-1}\equiv\frac{1}{8\pi M_{n_{*}-1}}\approx\frac{\sqrt{\pi\left(\alpha-2\pi\right)}}{2\pi\alpha}\label{eq: T H}
\end{equation}
are imaginary for $\alpha<2\pi$. Then, they are imaginary also for
$\alpha=1.49$ (isolated horizon framework \cite{key-14,key-15}),
$\alpha=2.46$ (Tanaka-Tamaki scenario \cite{key-16}) and $\alpha=4.44$
(Kong-Yoon scenario \cite{key-17,key-18,key-19}). This implies that
that transitions between two neighboring levels look forbidden for
highly excited black holes if one assumes the correctness of the analysis
in \cite{key-5}. Thus, in highly excited black holes, the minimum
energy found by Yoon \cite{key-5} should always correspond to transitions
between two levels which are not neighboring. 

Let us analyse this case in detail. A black hole excited at a level
$m$ has a mass \cite{key-9}

\begin{equation}
M_{m}\equiv M-(\omega_{0})_{m}\label{eq: M m}
\end{equation}
which, by using eq. (\ref{eq: radice fisica}) becomes

\begin{equation}
M_{m}=\sqrt{M^{2}-\frac{1}{4\pi}\sqrt{(\ln3)^{2}+4\pi^{2}m^{2}}}.\label{eq: M m 2}
\end{equation}
Considering two levels which are not neighboring, i.e. $m$ and $n$
with $n-m\geq2$, eq. (\ref{eq: radice fisica}) implies that one
needs the condition 

\begin{equation}
\begin{array}{c}
\sqrt{M^{2}-\frac{1}{4\pi}\sqrt{(\ln3)^{2}+4\pi^{2}m^{2}}}\approx\\
\\
\approx\sqrt{M^{2}-\frac{1}{4\pi}\sqrt{(\ln3)^{2}+4\pi^{2}n^{2}}}+\frac{\alpha}{8\pi\sqrt{M^{2}-\frac{1}{4\pi}\sqrt{(\ln3)^{2}+4\pi^{2}m^{2}}}}.
\end{array}\label{eq: bilanciamento-2}
\end{equation}
For $m,n\gg1$ eq. (\ref{eq: bilanciamento-2}) is well approximated
by 
\begin{equation}
\begin{array}{c}
\sqrt{M^{2}-\frac{1}{2}m}\approx\\
\\
\approx\sqrt{M^{2}-\frac{1}{2}n}+\frac{\alpha}{8\pi\sqrt{M^{2}-\frac{1}{2}m}}.
\end{array}\label{eq: bilanciamento-1-1}
\end{equation}
Putting 
\begin{equation}
\begin{array}{c}
M^{2}-\frac{1}{2}m\equiv x\\
\\
M^{2}-\frac{1}{2}n\equiv y\\
\\
\frac{\alpha}{8\pi}\equiv\beta,\;\; x\geq0,\;\; y\geq0
\end{array}\label{eq: putting-1}
\end{equation}
eq. (\ref{eq: bilanciamento-1-1}) becomes

\begin{equation}
\sqrt{x}\approx\sqrt{y}+\frac{\beta}{\sqrt{x}}\label{eq: bilanciamo-1}
\end{equation}
which can be simplified as

\begin{equation}
x-\beta\approx\sqrt{y}\sqrt{x}.\label{eq: bilanciamo 2-1}
\end{equation}
By squaring eq. (\ref{eq: bilanciamo 2-1}) and by dividing for $x$
one gets

\begin{equation}
y\approx x+\frac{\beta^{2}}{x}-2\beta.\label{eq: y di x}
\end{equation}
By using eqs. (\ref{eq: putting-1}) the condition $n-m\geq2$ implies
also $x-y\gtrsim1$, which, in turn, gives

\begin{equation}
2\beta-\frac{\beta^{2}}{x}\gtrsim1.\label{eq: proibire}
\end{equation}
Eq. (\ref{eq: proibire}) is easily solved for $x$:

\begin{equation}
x\lesssim\frac{\beta^{2}}{2\beta-1}=\frac{\alpha^{2}}{16\pi\left(\alpha-4\pi\right)},\label{eq: consistente}
\end{equation}
which is not consistent with the third of eqs. (\ref{eq: putting-1})
because the quantity $\frac{\alpha^{2}}{16\pi\left(\alpha-4\pi\right)}$
is always negative for $\alpha=1.49$ (isolated horizon framework
\cite{key-14,key-15}), $\alpha=2.46$ (Tanaka-Tamaki scenario \cite{key-16})
and $\alpha=4.44$ (Kong-Yoon scenario \cite{key-17,key-18,key-19}).
In other words, the mass (\ref{eq: M m 2}) results imaginary in this
case too.

Hence, in this work we have shown that the results by Yoon \cite{key-5},
which arise from loop quantum gravity, are not consistent with our
semi-classical results for highly excited black holes. Maybe the results
in \cite{key-5} can be correct for non-highly excited black holes,
but, in any case, our analysis renders further problematical the match
between loop quantum gravity and semi-classical theory.

\section*{Acknowledgements }

J.Mª Fernández Cristóbal has also to be thanked for correcting an
algebra error in the text. It is a pleasure to thank Hossein Hendi,
Erasmo Recami, Lawrence Crowell, Herman Mosquera Cuesta, Ram Vishwakarma
and my students Reza Katebi and Nathan Schmidt for useful discussions
on black hole physics.


\begin{thebibliography}{10}
\bibitem[1]{key-1}J. D. Bekenstein and V. F. Mukhanov, Phys. Lett.
B 360, 7 (1995).

\bibitem[2]{key-2}S. W. Hawking, Commun. Math. Phys. 43, 199 (1975).

\bibitem[3]{key-3}M. Barreira, M. Carfora and C. Rovelli, Gen. Rel.
Grav. 28, 1293 (1996, Second Award at Gravity Research Foundation).

\bibitem[4]{key-4}K. V. Krasnov, Class. Quant. Grav. 16, 563 (1999).

\bibitem[5]{key-5}Y. Yoon, arXiv:1210.8355 (2012).

\bibitem[6]{key-6}M. K. Parikh, Gen. Rel. Grav. 36, 2419 (2004, First
Award at Gravity Research Foundation).

\bibitem[7]{key-7}M. K. Parikh and F. Wilczek, Phys. Rev. Lett. 85,
5042 (2000).

\bibitem[8]{key-8}S. W. Hawking, Phys. Rev. D 14, 2460 (1976).

\bibitem[9]{key-9}C. Corda, Int. Journ. Mod. Phys. D 21, 1242023
(2012, Honorable Mention at Gravity Research Foundation).

\bibitem[10]{key-10}L. Motl, Adv. Theor. Math. Phys. 6, 1135 (2003).

\bibitem[11]{key-11}S. Hod, Phys. Rev. Lett. 81 4293 (1998).

\bibitem[12]{key-12}S. Hod, Gen. Rel. Grav. 31, 1639 (1999, Fifth
Award at Gravity Research Foundation).

\bibitem[13]{key-13}M. Maggiore, Phys. Rev. Lett. 100, 141301 (2008).

\bibitem[14]{key-14}A. Ashtekar, J. C. Baez and K. Krasnov, Adv.
Theor. Math. Phys. 4, 1 (2000).

\bibitem[15]{key-15}A. Ghosh and P. Mitra, 

\bibitem[16]{key-16}T. Tanaka and T. Tamaki, arXiv:0808.4056 {[}hep-th{]}
(2008). 

\bibitem[17]{key-17}B. Kong and Y. Yoon, arXiv:0910.2755 {[}gr-qc{]}
(2009). 

\bibitem[18]{key-18}B. Kong and Y. Yoon, arXiv:1003.3367 {[}gr-qc{]}
(2010). 

\bibitem[19]{key-19}B. Kong and Y. Yoon, to appear.

\bibitem[20]{key-20}C. Corda, Ann. Phys. 337, 49 (2013).

\bibitem[21]{key-21}C. Corda, Eur. Phys. J. C 73, 2665 (2013). 

\bibitem[22]{key-22}C. Corda, S. H. Hendi, R. Katebi, N. O. Schmidt,
JHEP 06 (2013) 008.\end{thebibliography}
\end{document}